\documentclass[useAMS,usenatbib]{mn2e}

\usepackage{graphics,epsfig}
\usepackage{graphicx}
\usepackage{amssymb}

\title[\HI\ in SBS 0335--052]{\HI\ in very metal-poor galaxies: the SBS~0335--052 system}
\author[Ekta et al.]
{B. Ekta,$^1$\thanks{ekta@ncra.tifr.res.in,sap@sao.ru,chengalu@ncra.tifr.res.in}
Simon A. Pustilnik,$^{2,3}$\footnotemark[1]
Jayaram N. Chengalur$^1$\footnotemark[1]  \\
$^1$ National Centre for Radio Astrophysics, Post Bag 3, Ganeshkhind, Pune
411 007, India\\
$^2$ Special Astrophysical Observatory of RAS, Nizhnij Arkhyz,
Karachai-Circassia 369167, Russia\\
$^3$ Isaac Newton Institute of Chile, SAO Branch, Nizhnij Arkhyz, Russia}

\DeclareRobustCommand{\ion}[2]{%
\relax\ifmmode
\ifx\testbx\f@series
{\mathbf{#1\,\mathsc{#2}}}\else
{\mathrm{#1\,\mathsc{#2}}}\fi
\else\textup{#1\,{\mdseries\textsc{#2}}}%
\fi}

\newcommand{\kms}{km~s$^{-1}$}
\newcommand{\HI}{\ion{H}{i}}
\newcommand{\HII}{\ion{H}{ii}}
\newcommand{\atoms}{atoms~cm$^{-2}$}
\newcommand{\sunn}{$_{\odot}$}

\begin{document}
 
\label{firstpage}

\date{Accepted 2009 April 29.  Received 2009 April 29; in original form 2009 March
4}

\pagerange{\pageref{firstpage}--\pageref{lastpage}} \pubyear{2009}

\maketitle

\begin{abstract}

We present Giant Metrewave Radio Telescope (GMRT), \HI\ 21cm observations of 
SBS~0335--052E and SBS~0335--052W, a close pair of dwarf galaxies, which are 
further unusual in being the most metal-poor 
star-forming galaxies known. We present images at several angular resolutions,
ranging from $\sim$40 to 4~arcsec. These images show that 
SBS~0335--052 is a strongly interacting system, with a faint diffuse \HI\
bridge seen at low resolution, and elongated tails seen at the higher
resolutions. The overall morphology suggests that the pair represents 
a major (as both galaxies have similar \HI\ masses) merger of extremely gas-rich 
galaxies, which is currently past the first close encounter. The low-resolution  
velocity field is dominated by the velocity 
difference between the two galaxies and the velocity gradient along the 
tidal features. However, for SBS~0335--052W  at least, at high angular
resolution, one sees a central velocity field that could be associated 
with the spin of the original undisturbed disc. The two galaxies have very
similar \HI\ masses, but very different optical properties and current star
formation rates. A possible reason for this is the differing amounts of
tidally-induced star formation, because of the different spin orientations
of these interacting galaxies. The highest angular resolution \HI\ images show
that the ionized superbubble, identified by Thuan, Izotov $\&$ Lipovetsky (1997),  
in the Hubble Space Telescope ({\it HST}) images of SBS~0335--052E, is extended 
along one of the diffuse tidal 
features, and that there is a high-density \HI\ clump at the other end of the
superbubble. The star formation in SBS~0335--052E occurs mainly in a group
of superstar clusters (SSCs) with a clear age gradient; the age decreases as
one approaches the dense \HI\ clump. We suggest that this propagating star
formation is driven by the superbubble expanding into a medium with a 
tidally-produced density gradient. The high pressures associated with the 
compressed material would also naturally explain why current star formation
is mainly concentrated in superstar clusters.

\end{abstract}


\section{INTRODUCTION}
\label{sec:intro}

    Low-mass star-forming galaxies with metallicties of Z\sunn/10 and
lower\footnote[1]{Throughout this paper, we use the new scale for solar 
metallicity, viz., Z\sunn\ corresponds to 12~+~$\log$~(O/H)~=~8.66 (\citet{Solar04}).}  
(i.e., 12~+~$\log$~(O/H)$~\leq~$7.65; the so called `eXtremely Metal-Deficient' (XMD) galaxies)   
are very rare in the local Universe (e.g., \citet{kunth2000}). 
According to \citet{pustilnik03}, they comprise less than 2~per~cent of all known
emission-line galaxies. No more than half a dozen (of the tens of thousands
of emission-line galaxies) are known to have metallicities near the bottom of 
the metallicity-distribution (Z$\sim$Z\sunn/35--Z\sunn/25). In this paper, we present 
GMRT \HI\ observations of the SBS~0335--052 system, which contains the lowest
known metallicity XMD galaxies. SBS~0335--052 consists of a galaxy pair,
SBS~0335--052E and SBS~0335--052W, with a projected separation of $\sim$22~kpc
(Pustilnik et al. 2001). Both galaxies, in the pair, are extremely metal-poor, 
and have oxygen abundances, 12~+~$\log$~(O/H) $\sim$7.29 
(SBS~0335--052E, \cite{izotov1997} and references therein) and 7.12 (SBS~0335--052W,
Izotov, Thuan $\&$ Guseva (2005) and references therein), respectively. The galaxies are
also peculiar in that they have very blue colours in their outer parts 
(indicative of small ages even for the older stellar population, e.g.,
\citet{papaderos1998} and 
\cite{pustilnik2004}), and have the bulk of their baryonic material in the form of
gas.  

   Observations of local, low-metallicity galaxies allow one to gain insight
into the modes of star formation in primeval objects, where the physical
conditions (e.g., low metallicity, large gas fraction, small gravitational
potential, etc.) are likely to be similar. \HI\ observations of local XMD
galaxies have been presented by us (e.g., Nan\c {c}ay Radio Telescope data by 
\citet{pustilnik07}, and GMRT data by \citet{chengalur2006} and \citet{ekta2006}). 
In particular, Ekta, Chengalur, Pustilnik (2008) discuss the \HI\ distribution, 
kinematics and star formation in three of the six most metal-poor star-forming 
galaxies, viz., DDO~68, UGC~772 (SDSS~J0113+0052) and SDSS~J2104--0035.
In the case of  SBS~0335--052, Very Large Array (VLA) \HI\ 21cm observations have been presented
earlier in \cite{pustilnik2001}. Our current GMRT observations are both
deeper, and also cover a larger range of angular resolutions than the earlier
VLA ones. Throughout this paper, we assume a distance of 53.6~Mpc for
SBS~0335--052. At this distance, 1~arcsec corresponds to a linear distance of $\sim$260~pc.

\section{OBSERVATIONS AND DATA REDUCTION}
\label{sec:obs}

SBS~0335--052 was observed in five different runs, on 2004 November 28, 29, 
December 10, 11, 12, at the GMRT, with a total bandwidth of 2~MHz, and a channel
resolution of 15.6~kHz (which corresponds to a velocity resolution of
$\sim$3.3~\kms). The total on-source time was $\sim$20~h. The flux
calibrators, 3C~48 and 3C~147, and the phase calibrator, PKS~J0323+0534 were
observed at appropriate intervals.
Each of the data sets was reduced in classic Astronomical Image Processing 
System ({\small AIPS}). The data reduction
steps included flagging of the bad visibility points, calibrating for phase and
bandpass shape. The visibility data from the different runs were combined into 
one data set using the task {\small 'DBCON'}. Since the differential Doppler shift 
between the different runs is small compared to the channel width, no Doppler 
correction was applied. After continuum subtraction, each channel was 
imaged using task {\small 'IMAGR'}. The parameters of the GMRT observations are 
given in Table~\ref{tab:obspar}.

\begin{table}
\caption{Parameters of the GMRT observations}
\label{tab:obspar}
\begin{tabular}{ll}
\hline
Date of observations       & 2004 November 28, 29,  \\
                           & December 10, 11, 12 \\
Field center R.A.(2000)    & 03$^{h}$37$^{m}$44.0$^{s}$ \\
Field center Dec.(2000)    & --05$^{\degr}$02$^{\arcmin}$40.0$^{\arcsec}$ \\
Central Velocity (\kms)    & 4040 \\
Time on-source  (h)        & $\sim$20 \\
Number of channels         & 128 \\
Channel separation (\kms)  & $\sim$3.3 \\
Flux Calibrators           & 3C~48, 3C~147 \\
Phase Calibrator           & PKS~J0323+0534 \\
Resolution (arcsec$^{2}$) (root mean & 43~$\times$~39 (1.16) \\
square (rms) noise in mJy~Bm$^{-1}$) & 21~$\times$~19 (0.85) \\
          & 9~$\times$~9 (0.66) \\
                                & 6~$\times$~5 (0.55) \\
          & 3~$\times$~3 (0.51) \\
\hline
\end{tabular}
\end {table}

\section{\HI\ MORPHOLOGY AND KINEMATICS}
\label{sec:res}

\begin{figure*}
\includegraphics[width=18.0cm]{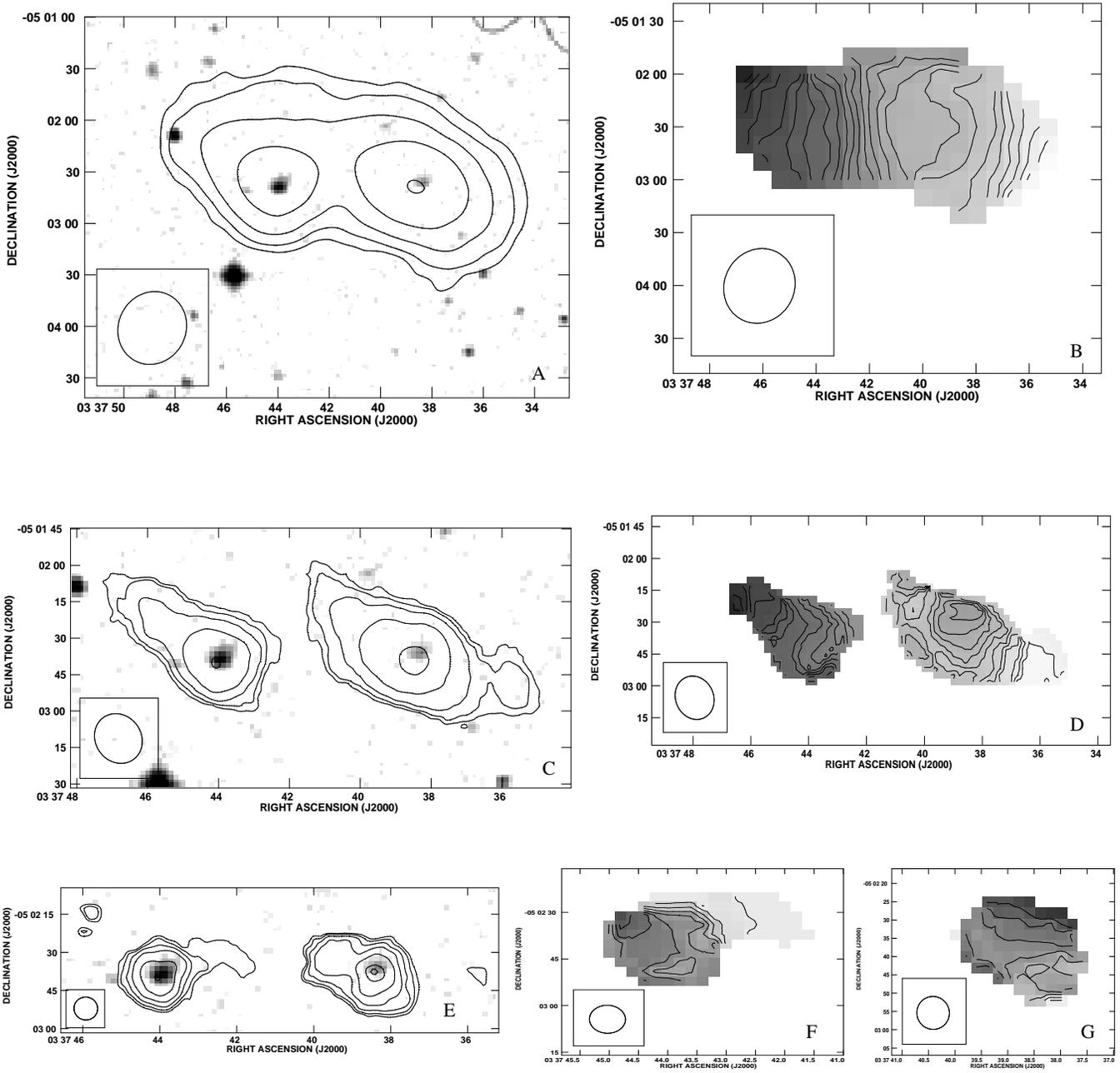}
\caption{
         {\bf (A.)} Integrated \HI\ emission map of SBS~0335--052 system,
     in contours, at a resolution of $\sim$43~$\times$~39~arcsec$^{2}$, 
     overlaid on the {\it B}-band Digitized Sky Survey-II (DSS-II) image (grey-scale, arbitrary units).
     The contour levels are at \HI\ column densities of $\sim$0.14, 0.31, 
     0.69, 1.52, 3.34~$\times$~10$^{20}$~\atoms. 
         {\bf (B.)} The \HI\ intensity-weighted velocity field (contours and
     grey-scales) at a resolution of $\sim$43~$\times$~39~arcsec$^{2}$.
     The velocity contours are from 3997 to 4074~\kms, in steps of of 3.3~\kms. 
         {\bf (C.)} Same as (A), excepting that the \HI\ image is at an angular resolution
      of $\sim$21~$\times$~19~arcsec$^{2}$. The contour levels are at \HI\ column
      densities of 0.30, 0.67, 1.50, 3.33 and 7.40~$\times$~10$^{20}$~\atoms. 
         {\bf (D.)} Same as (B), excepting that the \HI\ velocity field is at an angular 
      resolution of $\sim$21~$\times$~19~arcsec$^{2}$ resolution. The velocity 
      contours range from 3997 to 4080~\kms, in steps of 3.3~\kms.
         {\bf (E.)} Same as (A), excepting that the \HI\ image is at an angular resolution
      of $\sim$9~arcsec. The contour levels are at \HI\ column densities of
      0.75, 1.39, 2.57, 4.75, 8.78, 16.24 and 18.81~$\times$~10$^{20}$~\atoms.
         {\bf (F,G.)} Same as (B), excepting that the \HI\ velocity field is at an angular 
      resolution of $\sim$9~arcsec. In F, the velocity contours range from 4037 to 4060~\kms\
      (SBS~0335--052E), and in G, from 4013 to 4036~\kms\ (SBS~0334--052W). The contour
      spacing is 3.3~\kms.
}
\label{fig:mom}
\end{figure*}

The integrated \HI\ intensity map of SBS~0335--052 system, at an angular
resolution of $\sim$40~arcsec, is shown in Fig.~\ref{fig:mom}(A).
At this resolution, the two galaxies appear to be embedded in a common \HI\ envelope.
The earlier VLA observations (Pustilnik et al. 2001) also show the galaxies to be
in a common envelope, but the total \HI\ extent seen in our map 
(225~$\times$~105~arcsec$^{2}$, or 58.5~$\times$~27.3~kpc$^{2}$, at an
\HI\ column density level of 1.0~$\times$~10$^{19}$~\atoms) is smaller than
that deduced from the VLA observations (65~$\times$~22~kpc$^{2}$, after rescaling to
the distance adopted here). Our \HI\ maps are, at least, a factor of 1.5~times
more sensitive than those from earlier VLA observations. The rms noise quoted
by the latter is 1.0~mJy~beam$^{-1}$ per channel, for a channel width of
5.3~\kms, at an angular resolution of $\sim$20~$\times$~15~arcsec$^{2}$. 
Our maps, at slightly coarser resolution ($\sim$21~$\times$~19~arcsec$^{2}$), 
have an rms noise of 0.85~mJy~beam$^{-1}$ per channel, for a channel width of 3.3~\kms. 
The outermost contour, in Fig~\ref{fig:mom}(A), is a factor of
$\sim$5~times lower than the sensitivity level in the VLA data. The difference
in linear sizes seen at the GMRT and VLA is largely because a tentative feature 
(to the north-east of the main emission) seen in the VLA image has not been detected 
at the GMRT. 

\begin{figure}
\includegraphics[width=8.0cm]{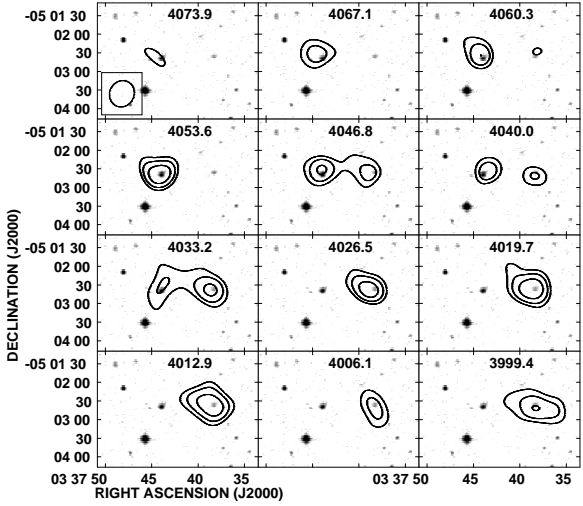}
\caption{The channel maps of SBS~0335--052 system, in contours, at 
a resolution of $\sim$43~$\times$~39~arcsec$^{2}$. The contours
are at -3, 3, 4.2, 6.0~times the rms noise in a single channel
(1.16~mJy~beam$^{-1}$). Every alternate channel is shown.
The {\it B}-band DSS-II image is shown in grey-scale, in arbitrary units. 
The velocity, for each shown channel, is given in corresponding panel.}
\label{fig:40arcchan}
\end{figure}

The GMRT \HI\ velocity field, at $\sim$40~arcsec, is shown in Fig.~\ref{fig:mom}(B). 
There is an overall velocity gradient in northeast--southwest (NE--SW) 
direction. The contours are crowded in the region between the two galaxies.  
A contour map (at the same angular resolution) of every alternate channel, 
in the data cube, is shown in Fig.~\ref{fig:40arcchan}. \HI\ emission, peaked 
at the location of each of the two galaxies, and an \HI\ `bridge', connecting 
the two galaxies (at intermediate velocities) can be seen. 

Fig.~\ref{fig:mom} also shows the \HI\ emission at two higher angular
resolutions, viz., $\sim$~20~arcsec (panels C, D) and $\sim$9~arcsec
(panels E, F, G). In the $\sim$20~arcsec image, the `bridge' emission
appears to be completely resolved out. To check if the `bridge' emission 
seen in panel~A is entirely due to beam smearing, we smoothed the
$\sim$20~arcsec map to $40$~arcsec resolution, and compared it with the map
shown in panel~A. The smoothed map does not recover all the flux seen in
the `bridge' region in panel~A, indicating that
the emission seen is not entirely due to beam smearing. At $\sim$20~arcsec 
resolution, one can clearly distinguish the \HI\ associated with each galaxy.
Both galaxies show extended elongated features, reminiscent of `tidal tails'.
The velocity field of the SBS~0335--052 system, at a resolution of
$\sim$21~$\times$~19~arcsec$^{2}$ in Fig.~\ref{fig:mom}(D), shows complex
patterns. Both galaxies have disturbed fields. In the case of SBS~0335--052E, 
there is a clear NE--SW velocity gradient, but the velocity contours show
substantial deviations from those expected from a rotating disc. In the 
case of SBS~0335--052W, there is an overall NE--SW velocity gradient along
the `tidal tails', but in the central part of the galaxy, the velocity
gradient is approximately from south to north. 

\begin{figure}
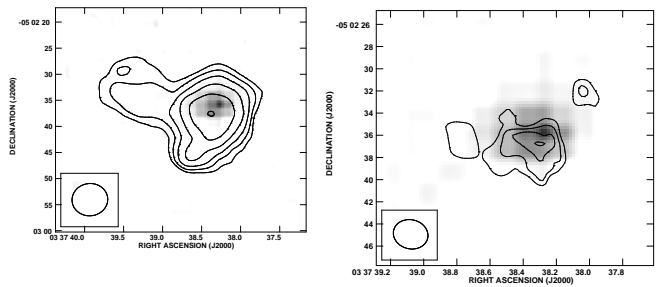

\includegraphics[width=4.05cm,angle=270]{SBS0335W6.5ARC}
\includegraphics[width=4.45cm,angle=270]{SBS0335W2.9ARC}
\caption{{\bf (Left.)} The integrated \HI\ intensity map of SBS~0335--052W, 
at an angular resolution of $\sim$7~$\times$~6~arcsec$^{2}$, in contours, overlaid on an DSS-II 
{\it B}-band image (grey-scale). The contours are at \HI\ column densities of
3.6, 5.3, 7.7, 11.4, 16.7 and 24.6~$\times$~10$^{20}$~\atoms.
The grey-scale is in arbitrary units. 
{\bf (Right.)} Same as on the left, except that the resolution of the \HI\ image is
$\sim$2.9~arcsec, and the contours are at \HI\ column
densities of 1.6, 2.3, 3.3 and 4.7~$\times$~10$^{21}$~\atoms.
}
\label{fig:hiresW}
\end{figure}

At a still higher angular resolution
(viz., $\sim$9~arcsec, panel E), one can see more clearly that the gas
distribution of each galaxy consists of a central peak coincident with the
optical emission, along with elongated `tidal tail' like features. In the
case of SBS~0335--052W, one can also see that the central \HI\ body is elongated in
the north-north-west--south-south-east (NNW--SSE) direction (see also Fig.~\ref{fig:hiresW}). 
The \HI\ velocity field, at this resolution (panel G), shows that the velocity gradient in
SBS~0335--052W is also approximately aligned with the major axis of the
central \HI\ distribution. This suggests that at this resolution one is
beginning to see the gas associated with the disc of this
galaxy. In the case of SBS~0335--052E, the dominant velocity gradient still
seems to be along the tidal tail. 

To summarize, our multi-resolution images
show SBS~0335-052 to be strongly interacting, with a diffuse \HI\ `bridge'
seen at low resolution, and elongated `tidal tails' seen at the higher
resolutions. The large-scale velocity field is largely dominated by the velocity of the
tidal features, however, for SBS0335--052W at least, at the highest angular
resolution, one sees a central velocity gradient that one could associate
with the spin of the original undisturbed disc.

\begin{figure}
\includegraphics[width=6.0cm,angle=270]{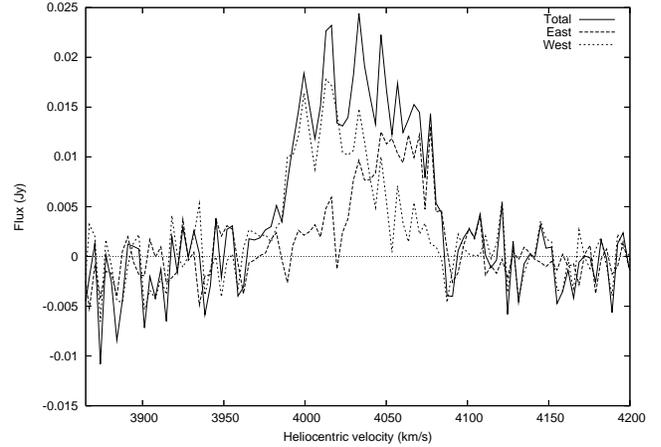}
\caption{The spectra of SBS~0335--052E and SBS~0335--052W (left) are overlaid
upon the spectrum of the whole system. These are obtained from data at an angular resolution of
$\sim$43~$\times$~39~arcsec$^{2}$.}
\label{fig:spectra}
\end{figure}

   The spectra of SBS~0335--052E, SBS~0335--052W and the whole system
(obtained from $\sim$40~arcsec data) are shown in Fig.~\ref{fig:spectra}.
The total integrated \HI\ flux is 1.46~$\pm$~0.15~Jy~\kms, which matches,
within error bars, with the value given in \cite{thuan1999}, viz., 
1.28~$\pm$~0.22~Jy~\kms. The VLA observations (Pustilnik et al. 2001) obtained a 
higher value of 2.46~$\pm$~0.18~Jy~\kms. The reason for this discrepancy (which is 
at the $\sim$4$\sigma$ level) is unclear.
We have confirmed the amplitude calibration of the GMRT maps, by comparing
the fluxes of background continuum sources in our map with their NRAO VLA Sky 
Survey (NVSS) fluxes.
The \HI\ fluxes measured for eastern and western components are 0.61 and
0.86~Jy~\kms\ (42 and 58~per~cent of the total flux), respectively,
corresponding to \HI\ masses of 4.2 and 5.8~$\times$~10$^{8}$~M\sunn, 
and an \HI\ mass-ratio of 1:1.4. The central velocities of SBS~0335--052E
and SBS~0335--052W are 4053.6~$\pm~$1.7, 4014.7~$\pm$~1.7~\kms, respectively,
and the velocity widths, at 50~per~cent levels, are 50.8~$\pm$~3.3 and 47.4~$\pm$~3.3~\kms. For the
entire system, the central velocity is 4031.5~$\pm$~1.7~\kms, the velocity
range over which emission is detected is  73~\kms, while the total linear
extent is $\sim$58.5~kpc. Assuming the system is bound, its indicative
dynamical  mass can be estimated as:

M$_{\rm ind}$~=~2.3~$\times$~10$^{5}$~$\times$~R$_{kpc}$~$\times$~V$^{2}_{km~s^{-1}}$~M\sunn
~=~$\sim$9.0~$\times$~10$^{9}$~M\sunn.

The same formula gives indicative dynamical masses of 6.0~$\times$~10$^9$~M\sunn\
and 7.2~$\times$~10$^9$M\sunn\ for the east (E) and west (W) galaxy, respectively (where we
have assumed sizes of 10 and 13.6~kpc, and velocity widths of 51
and 48~\kms, respectively).
The sum of the indicative dynamical masses of the two galaxies exceeds the total 
indicative dynamical mass, because the various inclination angles involved have 
not been correctly accounted for.

The indicative dynamical mass is $\sim$10 times larger than the total
gas mass.
Using the stellar masses from Pustilnik et al. (2004), the
M$_{\rm star}$/(M$_{\rm star}$ + M$_{\rm gas}$) ratio is 0.07 and 0.016 for
the E and W galaxy, respectively. The various measured
parameters for the two galaxies are summarized in Table~\ref{tab:mainpar}.

\begin{table}
\caption{Main parameters of the observed galaxies.} 
\label{tab:mainpar}
\begin{tabular}{lll}
\hline
 Parameter    & SBS~0335--052E & SBS~0335--052W\\
V$_{\rm hel}$ (\kms)  & 4014.7 & 4053.6 \\
12+$\log$(O/H) & 7.29 & 7.12 \\
\HI\ flux (Jy~\kms) &  0.61 & 0.86  \\
M$_{\rm HI}$ (10$^{8}$~M$_{\odot}$) & 4.2 & 5.8 \\
M$_{\rm star}$/(M$_{\rm star}$+M$_{\rm gas}$) & 0.07 & 0.016 \\
M$_{\rm dyn}$ (10$^{9}$~M$_{\odot}$)& 7.0 & 7.9 \\
\hline
\end{tabular}
\end{table}

\section{STAR FORMATION IN SBS~0335--052}
\label{sec:sf} 

\begin{figure}
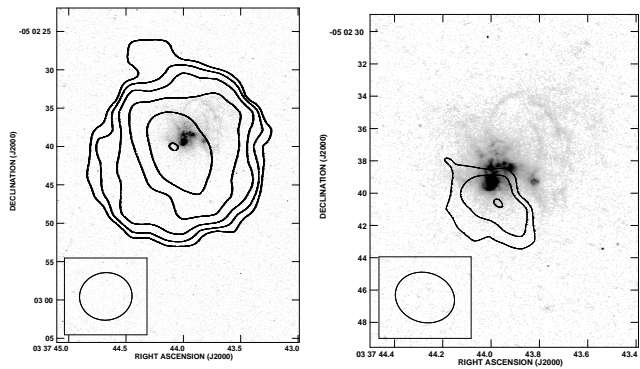

\includegraphics[width=4.05cm]{6.5arceast.eps}
\includegraphics[width=4.45cm]{3.4arcsbs0335east.eps}
\caption{{\bf (Left.)} The integrated \HI\ intensity map of SBS~0335--052E, 
at an angular resolution of $\sim$7~$\times$~6~arcsec$^{2}$, in contours,  
overlaid on an {\it HST} ACS continuum image (grey-scale). The contours are at
\HI\ column densities of 1.3, 2.3, 4.0, 6.9, 11.9,
20.6~$\times$~10$^{20}$~\atoms. The grey-scale is in arbitrary units.
{\bf (Right.)} Same as on the left, except that the resolution of the \HI\ image is
$\sim$4~$\times$~3~arcsec$^{2}$, and the contours are at \HI\ column
densities of 8.7, 13.2, 20.1~$\times$~10$^{20}$~\atoms.
}
\label{fig:hires}
\end{figure}

  Current star formation in SBS~0335--052E is largely confined to several
superstar clusters (SSCs), located in the central region of the galaxy
(\cite{thuanetal1997}). The ages of the SSCs show a systematic variation with
position, with the oldest SSCs being to the north and the youngest clusters
being at the southern tip (\cite{thuanetal1997,reines2008,thompson2009}).
Reines et al. (2008) compute an age of $\sim$15~Myr for the oldest SSCs and 
$<$3~Myr for the youngest ones, and also determine that the star formation
is propagating at a rate of $\sim$35~\kms. Thuan et al. (1997) also draw
attention to a large ($\sim$380~pc) superbubble, which they identify as being
a supernova blown cavity. Fig.~\ref{fig:hires}(left) shows the \HI\ emission
(contours) at resolution of $\sim$7~arcsec (1.8~kpc) overlaid on an {\it HST} 
Advanced Camera for Surveys (ACS) image. As can be seen, the \HI\ contours in 
the north-west are aligned with
the direction of the superbubble. The asymmetry of ionized gas, relative to
the central starburst, is also seen on the data from \citet{papaderos1998},
Pustilnik et al. (2004), \citet{izotov2006}, and recent Fabry-Perot interferometry 
to study H$\alpha$-kinematics (Pustilnik et al., in preparation). 
Note also that the north-west extensions of the \HI\ contours are
essentially due to material being pulled out to form the `bridge' between
SBS~0335--052E and W (see the lower resolution images, Fig.~\ref{fig:mom}).
At a still higher resolution of $\sim$4~arcsec (1~kpc,
Fig.~\ref{fig:hires}(right)), one can see that the detected \HI\ emission is
confined to a region to the south-east of the southern most SSC. The
sensitivity of the \HI\ observations decreases with increasing spatial
resolution, so the high resolution images detect only the highest column
density gas. As can be seen from Fig.~\ref{fig:hires}(left), the superbubble region 
corresponds to gas with column densities $\lesssim$9~$\times$~10$^{20}$~\atoms,
while the density in the south-east ridge (Fig.~\ref{fig:hires}(right)) reaches
up to $\sim$21~$\times$~10$^{20}$~\atoms. The superbubble seems to have expanded
preferentially into the lower density gas. The latter is produced when the `bridge'
material is being pulled out. The propagation of star formation towards
the south-east could be due to the propagation of the superbubble into the
relatively denser gas there.

In this context, it is interesting to note that the expansion
velocity that Thuan et al. (1997) derive for the superbubble is $\sim$17~\kms,
which is within a factor of $\sim$2 of that derived for the propagation
rate of the star formation. Given the large uncertainties in both these rates,
agreement to within a factor of 2 is quite reasonable. The north-to-south
propagation of the star formation, hence, appears to be attributable to the
expansion of a superbubble into a medium in which there is an overall north-south 
density gradient because of the tidal interaction. Such a scenario
would also provide a natural explanation for the dominant cluster mode of
star formation observed in SBS~0335--052E. High pressures are expected in gas
that is being compressed by an expanding supernova shell, and
\cite{elmegreen04} suggests that high pressure is conducive to the cluster
mode of star formation. A similar pattern of sequential star formation and
a dominant cluster mode for the star formation has recently been found
associated with an expanding supernova shell in the outer galaxy
\citep{kobayashi2008}.

Given that SBS~0335--052E and W are among the lowest metallicity galaxies known,
it is interesting to see if the star formation occurs at a similar gas
threshold density as for more metal-rich galaxies. For example,
\cite{skillman1987} suggested that star formation in dwarf galaxies
occurs only when the gas column-density crosses a threshold value of
$\sim$10$^{21}$~\atoms; this in turn could be related to a threshold amount
of dust shielding required for the production of molecular gas. In this case, 
one would expect that the threshold column-density increases with decreasing
gas-phase metallicity. \cite{schaye2004} (see also
\cite{schaye2008, schayed2008}) suggests a different model, one in which star
formation is related to the formation of a cold phase of the interstellar
medium. From his fitting formulae for the threshold column density, the
threshold density increases
with increasing gas fraction and decreasing metallicity (all other factors
being equal). In this model too, gas-rich XMD galaxies should have a higher
threshold for star formation than spirals. This issue was also explored, in
detail, in \cite{ekta2008},
who found no conclusive evidence for an increase in threshold density with
metallicity, albeit in a relatively small sample of XMD galaxies.

    The  \HI\ column densities at the positions of the brightest star-forming 
regions in SBS~0335--052E and SBS~0335--052W are $\sim$2.0 and
5.4~$\times$~10$^{21}$~\atoms, respectively. The measurements were made at
a resolution of $\sim$3.4 and 2.75~arcsec (880 and 715~pc), 
respectively. The signal-to-noise ratio for SBS~0335--052E was too low for
the column density to be measured at the higher resolution used for
SBS~0335--052W. In SBS~0335--052E, the \HI\ peak is slightly
(i.e., $\sim$1.5~arcsec, or 0.4~kpc), offset from the
brightest star-forming region (as was already noticed from the VLA data
(Pustilnik et al. 2001)). Nevertheless, it lies within a half beam-width of
the latter. The western \HI\ peak is also slightly ($\sim$1~arcsec, or
0.26~kpc) offset from the brightest star-forming region in
SBS~0335--052W. Note that the column densities have not been corrected for
inclination, because, as discussed above, it is difficult to estimate the
inclination angles of the \HI\ discs. The \HI\ column density of
SBS~0335--052E, as calculated from its {\it HST} spectral  observations
by \cite{thuan1997}, is 7.0~$\pm$~0.5~$\times$~10$^{21}$~\atoms. The higher
value that they obtain is presumably a result of the much higher angular
resolution of the {\it HST} data. These peak \HI\ column densities, are amongst
the largest values found in gas-rich dwarf or XMD galaxies (see, e.g., 
\cite{taylor94,vanZee1998,begum2006}; Ekta et al. 2006; Ekta et al. 2008). This is despite
the fact that all the above mentioned studies have better linear resolution 
than that of our observations, except for a few galaxies in Taylor et al. (1994).
Given the uncertain inclination correction however, it is difficult to 
gauge the significance of this higher observed \HI\ column density. 

\section{DISCUSSION AND SUMMARY}
\label{sec:dis}

\begin{figure*}
\begin{center}
\hskip -8.0cm \includegraphics[height=6.0cm,angle=-90]{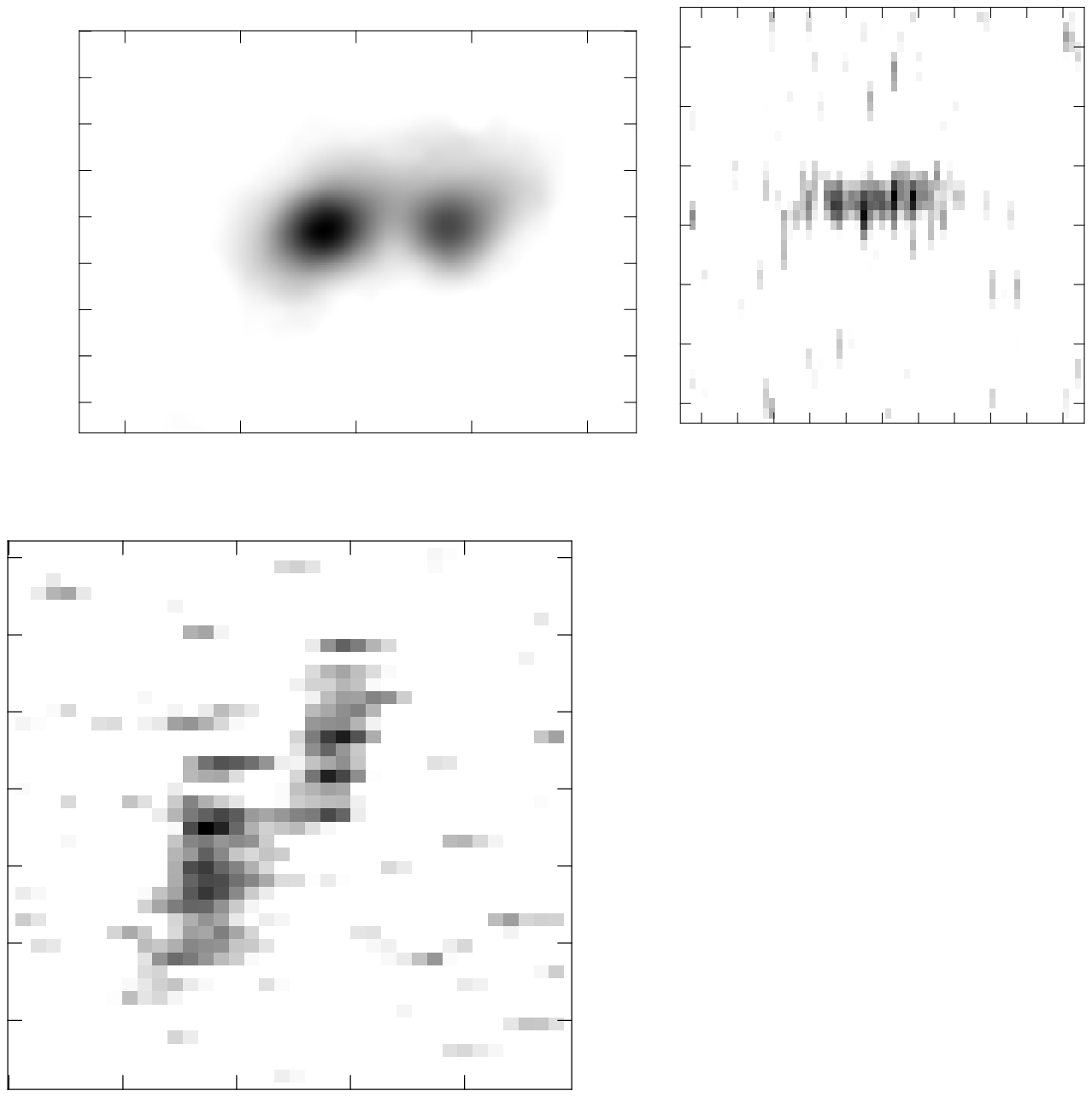}
\vskip -6.0cm \hskip 8.0cm \includegraphics[height=6.0cm]{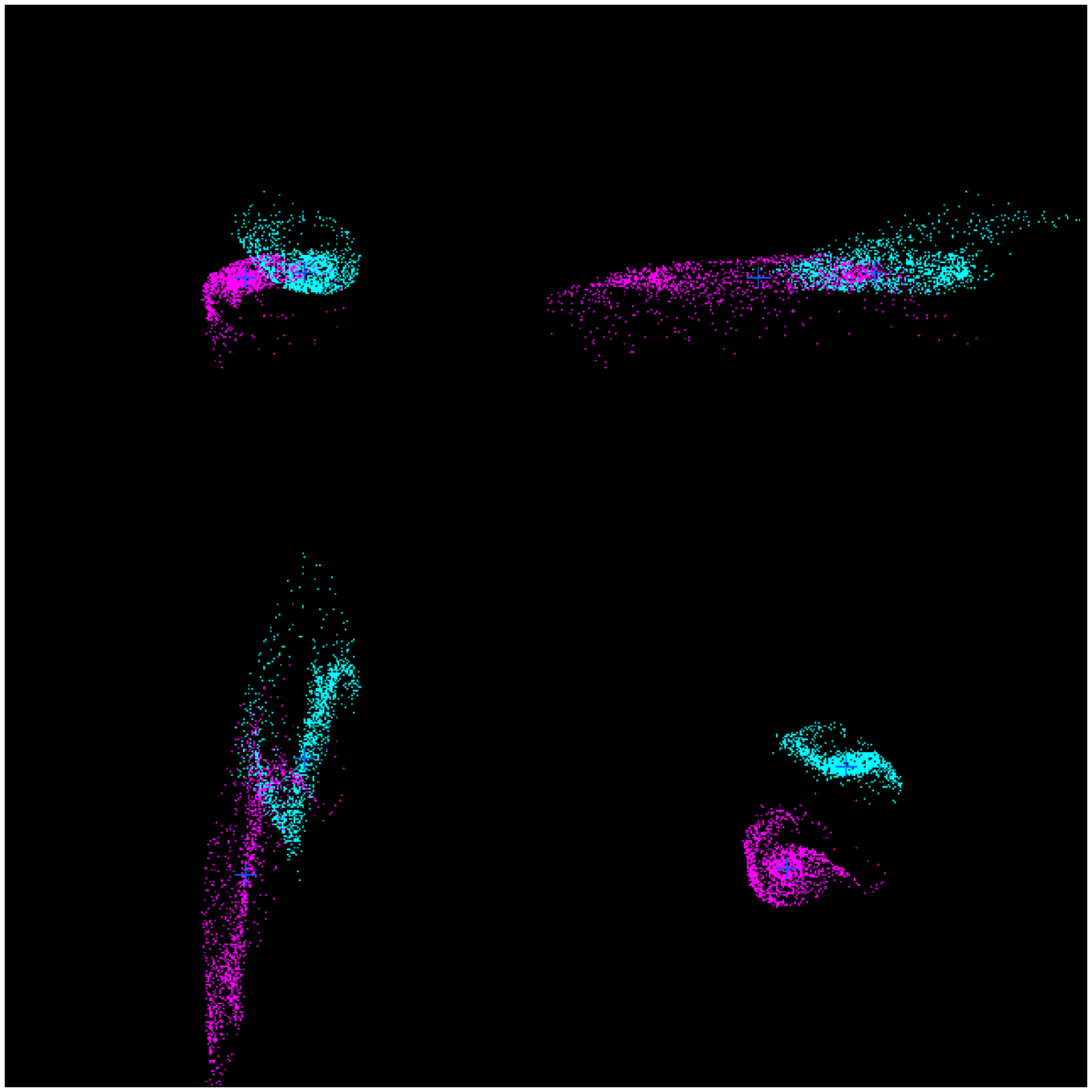} \hfill
\end{center}
\caption{
{\bf Left:} Projections of the \HI\ data cube, made for comparison with the
Identikit model. The panels (clockwise and starting from the upper left)
are {\bf (1)}~The sky ($X,Y$) plane projection of the \HI\ in SBS~0335--052
system, at an angular resolution of $\sim$40~arcsec. Note that the RA (`X')
axis is flipped with respect to Fig.~\ref{fig:mom}, so the E galaxy is to the
right. {\bf (2)}~The $V,Y$ plane projection (where `V' is the velocity). (3)~The
$X,V$ plane projection.
{\bf Right:} Identikit web interface results for the closest fit model that
we could make for our data. The panels are in the same order as for the
observed \HI\ data. The lower right panel is the $X,Z$ plane projection,
where $Z$ is the depth along the line of sight (for which there is no
counterpart in the left panel, since the $Z$ co-ordinate is not an
observable). Blue dots correspond to the first galaxy, and pink ones correspond to the
second galaxy. The input parameters used in the model are pericenter distance
$p$ = 0.125 (1.5~$h$), inclinations of galaxy spins relative to the orbital
plane $i_{1}$=130, $i_{2}$=70~degrees, azimuthal orientations
$\omega_{1}$=340, $\omega_{2}$=90~degrees, the angles between the sky plane
and the orbital plane, $\theta_X$=45 and $\theta_Y$=45~degrees.
The rotation angle around $Z$-axis is 310~degrees. The snapshot is for a time
$t \sim$0.5 (or $\sim$70~Myr, in physical units), after the first close passage.
The conversion from dimensionless units to physical units was done assuming
that the galaxies in SBS~0335--052 system have scale-length, $h \sim$0.37~kpc and, 
V$_{\rm rot} \sim$40~\kms.
}
\label{fig:identikit}
\end{figure*}

SBS~0335--052 is the lowest metallicity star-forming system known. The system
is further peculiar in being an interacting pair of dwarf galaxies. In the
case of SBS~0335--052E, the metallicity is low not just in the
ionized \HII\ regions, but also in the surrounding \HI\ gas. From FUSE 
observations, \cite{thuan2005} find that the metallicity in the \HI\ 
envelope is similar to that in the central star-forming regions. 

  Our \HI\ data shows that the pair is in an advanced stage 
of interaction, with the gas being pulled out to form a bridge, as 
well as into elongated tidal tails. As discussed in Section~\ref{sec:res},
if the dynamic-to-baryonic mass ratio is $\gtrsim$10, the system
is bound, and will eventually merge. The relatively short tails, and the
faintness of the `bridge' then suggest that the pair is probably past its
first close encounter, but has yet to fall towards each other for the second
time \citep[see, e.g.,][]{toomre72}. \cite{dimatteo08} show that starbursts
can be triggered in extremely gas-rich galaxies, which are undergoing
their first close encounter, depending on the details of the galaxy structure
and the orbit geometry. The starburst onset could range from being just before
the first close encounter (e.g., for very close, planar encounters) to
0.1--0.4~Gyr after the encounter. The relatively broad tidal features
(as opposed to the narrow long features produced by direct encounters) also
suggest that neither of the galaxies is undergoing a direct encounter
\citep[see, e.g.,][]{howard93}. Indeed from Fig.~\ref{fig:mom}, 
one can see that if the elongated structure seen in the central
regions of SBS~0335--052W reflects the underlying gas disc, the galaxy
is undergoing a highly inclined encounter.

Interestingly, all of the six lowest-metallicity star-forming XMD galaxies known are
either currently
undergoing tidal interactions, or appear to be the remnants of a merger
of extremely gas-rich progenitors \citep{ekta2008}. While interactions
have long been regarded as triggers for starbursts, numerical simulations
show that, in fact, only a small fraction of all binary interactions lead to
a reasonably strong starburst \citep{dimatteo08}. The enhancement
in the star formation rate during merger depends on a number of factors,
but in general (in recent  models) the internal structure (e.g.,
presence of a bulge, gas-mass fraction, disc-stability factor, etc.)
appear to play a more crucial role in determining the 
strength and duration of the starburst than the interaction geometry
\citep{mihos96,dimatteo08}. In particular, for gas-rich galaxies,  
\cite{dimatteo08} find that the amplitude of the starburst, which 
occurs just after the first close passage, is generally stronger 
than for gas-poor galaxies.

In the current situation, the galaxies concerned are both extremely gas-rich
dwarfs, and are likely to have similar internal structure. While numerical 
simulations of such systems are very limited, for somewhat less gas-rich
systems, \cite{mihos96} show that coplanar orbits lead to somewhat 
stronger starbursts than inclined orbits. As discussed above, SBS~0335--052W
appears to be undergoing an inclined encounter, which could perhaps
be the reason for its relatively lower star formation rate. As far
as simulations of interacting, gas-rich galaxies are concerned,
\cite{springel05} show that a direct encounter between two giant, 
extremely gas-rich discs can lead to the formation of spiral-galaxy-like
gas-rich remnant. The star formation rate is enhanced both during
the first close passage as well as during the final merger. In general,
gas-rich mergers tend to show larger enhancements in the instantaneous
star formation rate than gas-poor ones, both because there is more raw
material available for star formation, and also because of the inherent
instability of thin gas discs \citep{dimatteo08}.

     We used the recently released Identikit tool \citep{barnes09} to 
model the interaction in the SBS~0335--052 system. Since Identikit is based
on collisionless N-body simulations, we compare only the morphology of the 
outer regions and the velocity-field projections. The best possible
match, that we could find, is shown in Fig.~\ref{fig:identikit}, along
with the data projections in the same co-ordinate system. While the match 
is not perfect (given the limitations of the model, a perfect match
is perhaps not to be expected) the parameters from the match could
be taken as indicative. In the displayed model, the galaxies undergo
a close (pericenter distance of $\sim$0.55~kpc) encounter, which is
somewhat retrograde encounter for SBS0335--052E, and close to polar
for SBS~0335--052W. The sky plane is inclined $\sim$45~degrees with respect
to the orbit plane. The observed snapshot corresponds to shortly ($\sim$0.1~Gyr) 
after the first perigalacticon. From optical colours,
the major enhanced star formation in both components
is estimated to have started  0.1--0.4~Gyr ago. It is interesting to note 
that this is consistent with the star formation acitivity starting just before 
the pericenter passage, as predicted for some cases of very close 
encounters in gas-rich mergers \citep{dimatteo08}. A more detailed
simulation of this unique pair would be highly desirable.

While the interaction in SBS~0335--052 is likely to have triggered the current
starburst, it almost certainly plays an important role in the long-term
evolution of the system. \cite{bekki08} presents
results from numerical simulations of gas-rich dwarf galaxies, and shows
that in the case where there is no stellar bulge, and the gas is initially
in a cold disc, the merger remnant resembles a moderately gas-rich BCD (blue compact dwarf) galaxy.
The central compact emission comes from a starburst triggered by the
inflow of gas from the outskirts of the initial cold disc. High gas 
pressures in the star-forming regions also lead to the formation
of massive star clusters (as seen for SBS~0335--052E). As mentioned
above, the role of high pressures in forming massive star clusters
was also discussed earlier by \cite{elmegreen1993} and \cite{elmegreen04}. 
Further, \cite{dimatteo08} find that the
strength of the central starburst found in merging gas-rich galaxies
depends critically on the assumed disc stability. Simulations in which
gas dissipation has been decreased to prevent excessive disc fragmentation
for isolated galaxies (e.g., their particle-mesh - sticky-particles (PM-SM) simulation) show very little
enhancement of star formation during the merger. Further, even in their
simulations which do show an enhancement in the star formation rate, the
major cause of the enhancement is induced disc instabilities, and not inflow
of gas. While a significant radial metallicity gradient is known to exist 
in large spiral galaxies, such strong gradients have not been found in dwarf galaxies, 
and hence, the role that infalling gas plays in lowering the metallicity of the central
regions may not be important. 

Inflow of metal-poor gas is one possible cause for sizable metal
deficiency in central star-forming regions of some dwarf galaxies.
Other possible causes are (a) preferential loss of metal-enriched gas, i.e., 
escape of supernova metal-enriched material from previous starbursts,
(b) infall of pristine `gas clouds' from the intergalactic medium, (c)
very slow evolution of locally very stable gas discs (extreme low surface brightness (LSB) galaxies),
which fortuitously (e.g., due to residing in a void) escaped strong external
disturbances and related additional star formation,
and (d) a genuinely young galaxy which has not undergone much star formation
(i.e., due to a too small an age).

Numerical simulations \citep[e.g.,][]{fragile} show that sufficiently strong
starbursts, in sufficiently small galaxies, could indeed lead
to significant outflow of metal-enriched gas. A high spatial and velocity
resolution study of the ionized gas around the starburst region in 
SBS~0335--052E has indeed identified a high velocity ($\sim$50~\kms)
outflow \citep{izotov2006}. The final fate of this outflowing gas (i.e.,
whether it escapes from the galaxy or falls back to form a `galactic
fountain') depends on several parameters, including the mass of the parent
galaxy, the presence of a hot-gas halo from the previous star formation activity, etc.. In
addition, the positive feedback from massive, compact superstar clusters
could also inhibit outflow of gas \citep[e.g.,][]{tenorio2005,tenorio2007, 
wunsch2008}. As far as infall of pristine gas clouds is concerned, our
present data does not show any evidence for this, although the possibility 
that higher sensitivity data may show such clouds cannot be ruled out.

The SBS~0335--052 system is especially unusual in that both galaxies in the
pair are extremely metal-deficient, with metallicities differing by
only a factor $\sim$1.5. One possible explanation for their similar
metallicities is that there has been an exchange and mixing of gas during
their interaction. However, this seems unlikely, because of the faintness
of the observed `bridge', and also because gas exchange is not believed to
be very efficient during the first encounter of equal-mass galaxies 
\citep{toomre72}. It seems likely therefore, that the similar metallicities
of these two galaxies are due to similar evolutionary histories. In this
context, it is interesting to note that the SBS~0335--052 system is located
near the edge of a large void \citep{peebles01}, and that there appears to
be, in general, an indication of excess of metal-deficient galaxies in
voids \citep{pustilnik06,pustilnik07a}.

It is also interesting to note that the
most metal-poor galaxies known, viz., SBS~0335--052, I~Zw~18 and DDO~68, all
appear to be interacting or merger remnants. Indeed, in the order given above,
one could take these galaxies as forming a `merger sequence',
(with SBS~0335--052 being just past the first close passage, DDO~68 being a
merger remnant \citep{ekta2008,pustilnik08} and I~Zw~18 \citep{vanZee1998}
being at an intermediate phase) of extremely gas-rich galaxies.

   To summarize, our multi-resolution \HI\ observations show that
the SBS~0335--052 system consists of a merging galaxy pair. At the lowest
resolution, we see a faint diffuse `bridge' joining the two galaxies
(although the possibility of all of the observed emission arising from
beam smearing can not be completely ruled out), while
at higher resolutions, we see both \HI\ concentrations around the
stellar bodies as well as elongated tidal tails. The velocity fields
are complex, and, at large-scales are dominated by the velocity difference
between the galaxies and the velocity gradient along the tidal tails. At 
still higher resolution, we see that in SBS~0335--052E, the ionized 
superbubble identified in the {\it HST} images by Thuan et al. (1997) is extended
along one of the tidal tails, while there is a clump of high-density \HI\ 
gas at the other end of the superbubble. This is consistent 
with a scenario, in which the propagating star formation seen in this
galaxy is driven by a superbubble expanding into a medium with a 
tidally-produced density gradient. In contrast to SBS~0335--052E, the
higher resolution maps for SBS~0335--052W  suggest that it is
undergoing a somewhat polar encounter. Given the similar \HI\ parameters 
of both galaxies, this difference in encounter geometry
is possibly the reason for the weaker tidal disturbance and the
resulting large difference in the past and current star formation
in the E and W components.

\section*{AKNOWLEDGEMENTS}
We thank the staff of the GMRT who have made these observations possible.
The GMRT is run by the National Centre for Radio Astrophysics of the Tata
Institute of Fundamental Research.
Partial support for this work was provided by ILTP grant B-3.13. SAP
appreciates the partial support of this work through the Russian Foundation
for Basic Research (RFBR) grant 06--02--16617 
and the Russian Federal Agency on Education, project code 2.1.1/1937.
The authors appreciate the assistance of the `Identikit' authors J.~Barnes
and J.~Hibbard.

\label{lastpage}

\end{document}